\documentstyle[]{article}
\begin{document}
\begin{center} {\Large {\bf QED in optically active media:
Enhanced spontaneous emission and
chiral yet parity conserving corrections to the electron
mass}}\\[1cm]
Karl-Peter Marzlin
\footnote{e-mail: peter.marzlin@uni-konstanz.de}
\\[2mm]
Fakult\"at f\"ur Physik
der Universit\"at Konstanz\\
Postfach 5560 M 674\\
D-78434 Konstanz, Germany
\end{center}
$ $\\[3mm]
\begin{minipage}{15cm}
\begin{abstract}
The electromagnetic field inside an isotropic optically active
medium is quantized on the base of Fedorov's model for optical
activity. The modified photon propagator is derived. Using this
result it is shown that the QED correction to the electron mass
contains chiral terms which nevertheless are parity invariant.
The spontaneous emission rate of a two-level atom is shown to
increase in the medium. The spontaneously
emitted radiation is partially polarized.
\end{abstract}
\end{minipage}
$ $ \\[5mm]
PACS: 12.20.-m,33.55.Ad
\\[3mm]
{\bf Introduction}\\
In recent years there has been a growing attention to quantum
electrodynamics (QED) in dielectric media
\cite{hillery84,drummond90,glauber91,barnett92,milonni95}. This
field is of interest for several reasons. One motivation is the
study of squeezed states of light connected with the dielectric
\cite{glauber91}.
Another one is the more accurate description of Casimir forces
between conducting plates which are special cases of dielectrics
(see, e.g., Ref. \cite{milonni94} and references therein).
Despite these efforts the case of optically active media,
for which (in most models) the electric displacement vector depends
also on derivatives of the electric field, was not often studied.
To the author's knowledge there is only a work of Eimerl
\cite{eimerl88} in which the photon self energy is used to
give predictions for the parameters describing the medium, and
a paper of Woolley \cite{woolley71} where a nonrelativistic QED
calculation for the molecular magnitudes leading to optical
activity is presented.

The influence of the quantized electric
field on quantum matter in the presence of an optically active
medium seems not to be discussed in the literature. In this paper
it is considered how mass renormalization for the
electron and spontaneous emission will be modified by the medium.
Compared to ordinary dielectrics new effects mainly arise
because of the different behaviour of left and right circularly
polarized field modes. Since these polarization modes are the
eigenmodes of the spin operator many new effects depending on the
spin and the angular momentum of atoms or electrons can be
imagined. This is the case for the self energy of the electron,
for instance, which will be shown to include chiral terms.
The spontaneous emission of photons, which in the
approximation made here does not include spin or angular
momentum of the atom, leads to partially polarized light as
will be demonstrated later.
Beside the examples studied here the Casimir effect
for optically active media is under investigation.
It should be mentioned, however, that the effects produced by
the medium are in general small since the change in the refractive
index in optically active media is small compared to the change
in ordinary dielectrics.\\[2mm]
{\bf Quantization of the electromagnetic field}\\
The starting point of the calculations are the Maxwell equations
\begin{equation} \begin{array}{ll} \nabla \cdot\vec{D} = \rho & ,
  \nabla \times\vec{H} - \dot{\vec{D}} = \vec{j} \\
  \nabla \cdot\vec{B} = 0 & ,
  \nabla \times\vec{E} + \dot{\vec{B}} = \vec{0} \end{array}  
\end{equation}
together with Fedorov's model for optical activity
\cite{fedorov59}
\begin{eqnarray} \vec{D} &=& \varepsilon ( \vec{E} + \beta
  \nabla \times \vec{E}) \nonumber \\
  \vec{B} &=& \mu ( \vec{H} + \beta
  \nabla \times \vec{H})\label{mateqns} \end{eqnarray}
where only the case of an isotropic medium is considered. For the
sake of simplicity $\varepsilon$ and $\mu$, which are not related
to the optical activity, will be replaced in this paper by their
vacuum value, $\varepsilon = \varepsilon_0$, $\mu = \mu_0$.
Fedorov's model was chosen to preserve gauge invariance
(for a compilation of other models see Ref. \cite{mcclain93}).
The reason for this is that the nontrivial part of the material
equations (\ref{mateqns}) depends on $\vec{E}$ and $\vec{B}$ only
through their curl. This implies that only the transverse
electromagnetic field, which is not affected by gauge
transformations, will be modified by the medium.

Switching to the gauge potentials $\phi , \vec{A}$
and imposing the Lorentz gauge
condition on them one easily deduces the field equations
\begin{eqnarray} \Box \phi &=& 0 \nonumber\\
   \Box \vec{A} &=& - \frac{\beta}{c^2} \partial_t^2 \{ 2
   \nabla \times \vec{A} + \beta \nabla \times \nabla \times
   \vec{A} \}\label{feldgln}\end{eqnarray}
in absence of free charges $\rho$ or sources $\vec{j}$. The
d'Alembertian $\Box$ is defined by $\partial_t^2 /c^2 - \Delta$.
To quantize these fields the method of normal mode decomposition
will be used. To do so one first shows that the scalar products
\begin{eqnarray}
  (\phi , \psi ) &:=& \frac{-i \varepsilon_0}{\hbar c^2} \int
  d^3 x \{ \dot{\phi}^\dagger \psi - \phi^\dagger \dot{\psi} \}
  \nonumber \\
  (\vec{A},\vec{A}^\prime ) &:=& \frac{-i \varepsilon_0}{\hbar}
  \int d^3 x \Big \{ \dot{\vec{A}}^\dagger \cdot \vec{A}^\prime -
  \vec{A}^\dagger \cdot \dot{\vec{A}}^\prime + 2 \beta \Big [
  (\nabla \times \dot{\vec{A}}^\dagger)\cdot \vec{A}^\prime -
  \vec{A}^\dagger \cdot (\nabla \times \dot{\vec{A}}^\prime)\Big ]
  + \nonumber \\ && \hspace{2cm} \beta^2 \Big
  [(\nabla \times \dot{\vec{A}}^\dagger)\cdot (\nabla \times
  \vec{A}^\prime) - (\nabla \times \vec{A}^\dagger)\cdot
  (\nabla \times \dot{\vec{A}}^\prime)\Big ] \Big \} \end{eqnarray}
are conserved for solutions $\phi ,\psi$ and $\vec{A},
\vec{A}^\prime$ of Eqs. (\ref{feldgln}). The general solution of
the field equations can be found by Fourier transformation. It is
given by
\begin{eqnarray} \phi (x^\mu) &=& \sqrt{\frac{\hbar c}{(2\pi)^3
  \varepsilon_0}} \int \frac{d^3 k}{\sqrt{2k}} \Big \{ a_0(\vec{k})
  e^{-i k\cdot x} + h.c. \Big \} \label{lsg} \\
  \vec{A}(x^\mu) &=& \sqrt{\frac{\hbar}{(2\pi)^3
  \varepsilon_0 c}} \int d^3 k \Bigg \{ \vec{e}_3 (\vec{k})
  \frac{a_3(\vec{k})e^{-ik\cdot x}}{\sqrt{2k}} +\vec{e}_+(\vec{k})
  \frac{a_+(\vec{k})e^{-ik_+\cdot x}}{\sqrt{2k_+^0}(1-\beta k)}
  +\vec{e}_-(\vec{k})
  \frac{a_-(\vec{k})e^{-ik_-\cdot x}}{\sqrt{2k_-^0}(1+\beta k)}
  + h.c. \Bigg \} \nonumber \end{eqnarray}
Here $\vec{e}_i$ are the complex polarization vectors defined by
$\vec{e}_3 := \vec{k}/k$ and $i \vec{e}_3 \times \vec{e}_\pm =
\pm \vec{e}_\pm$. Although the calculation is not Lorentz invariant
it is convenient to adopt the covariant
notation $ a\cdot b := a^\mu b_\mu = a^0 b^0-\vec{a}\cdot \vec{b}$
and $x^0 := c t$. Throughout the paper the summation convention
over repeated indices is adopted. Greek indices are running from 0
to 3 and latin ones from 1 to 3. The various four-wavevectors
are defined by $ k^\mu := (k, \vec{k})$ and $ k^\mu_\pm
:= (k^0_\pm , \vec{k})$ where
\begin{equation} k^0_\pm := \frac{k}{\sqrt{1 \pm \beta k -
   \beta^2 k^2}} \; . \end{equation}

Here one can see the main effect of the presence of the optically
active medium. The dispersion relation between frequency and
wavevector of the left and right circularly polarized modes ($\pm$)
has been changed from $k^0 =k$ to $k^0 = k^0_\pm$. Since $k^0_\pm$
diverges for plus-modes with $\beta k = 1 + \sqrt{2}$ and
minus-modes with $\beta k = \sqrt{2}-1$ it is clear that the theory
describes physics only for electromagnetic waves with $\beta k \ll
1$. This should not be regarded as a shortcoming since in real
media optical activity appears only well below the high energy
physics regime. The introduction of a cutoff in momentum space
is therefore physically meaningful.

It should be mentioned that the dispersion relation $k^0 = k^0_\pm$
is a good description only for a certain part of the low energy
frequency range. This is a consequence of describing optical
activity within a local field theory. Despite this fact local
theories can still be a useful tool estimate the order of magnitude
of the medium's influence on quantum matter.

In Eq. (\ref{lsg}) it also becomes apparent that only the
transverse field modes ($\pm$) feel the presence of the medium in
Fedorov's model. Hence no gauge transformation can alter the
effect of the medium on the Maxwell field.

The quantization of the Maxwell field is now an easy task. One
demands the usual Lorentz gauge commutation relations
\begin{equation} [a_\alpha(\vec{k}), a_\beta^\dagger(\vec{k}^\prime
  )] = \delta (\vec{k}-\vec{k}^\prime) \delta_{\alpha \beta}
  \end{equation}
with $\alpha ,\beta = 0,+,-,3$ and $ -\delta_{00}=\delta_{33} =
\delta_{++}=\delta_{--} =1$ as well as $\delta_{\alpha \beta}=0$
for any other combination of $\alpha$ and $\beta$. In addition
one has to impose the Gupta-Bleuler condition $ \partial_\mu
A^{\mu (+)}|\psi \rangle =0 $ on the physical states $|\psi
\rangle$ (see, e.g., Ref. \cite{itzykson80}).\\[2mm]
{\bf The self energy of the electron}\\
To calculate the first order mass correction for the electron
the Feynman propagator
\begin{equation}
D^{\mu \nu}(x-y) = -i \theta (x^0-y^0) [A^{\mu (+)}(x), A^{\nu (-)}
  (y)] -i \theta (y^0-x^0) [A^{\nu (+)}(y), A^{\mu (-)}(x)]
\end{equation}
of the photon field is needed. From Eq. (\ref{lsg}) it is clear
that only the spatial part of the propagator is affected by the
medium. It is useful to rewrite $D^{ij}$ into the form $D^{ij
}_{vac} + D^{ij}_+ + D^{ij}_-$
where $D^{ij}_{vac}$ stands for the vacuum contribution and
$D^{ij}_\pm$ describes the changes of the propagation of the
left and right circularly polarized modes induced by the medium.
Using standard methods (including an integration over $k^0$ with
the residue formula and the use of $\vec{e}_+^i (-\vec{k})
\vec{e}_+^{j*} (-\vec{k}) = \vec{e}_+^{i*} (\vec{k}) \vec{e}_+^{j}
(\vec{k})$ and a similar equation for $\vec{e}_-$) the expression
for $D^{ij}_\pm$ in momentum space is found to be
\begin{equation}
D^{ij}_\pm (k^\mu) = \frac{\hbar}{\varepsilon_0 c} \vec{e}_\pm^i
  (\vec{k}) \vec{e}_\pm^{j*}(\vec{k}) \Big \{ \frac{1}{(1\mp
  \beta k)^2[(k^0)^2- (k^0_\pm)^2 +i \varepsilon]} -
  \frac{1}{(k^0)^2 - k^2 + i \varepsilon} \Big \}\; . \label{dij}
\end{equation}
Having found the photon propagator one is ready to calculate the
the first order on-shell mass correction \cite{itzykson80}
\begin{equation}
\delta m = \frac{i e^2}{(2\pi)^4 \hbar c} \int d^4 k D^{\alpha
  \beta}(k) \gamma_{\alpha} S_F(q-k) \gamma_{\beta}
\end{equation}
with $S_F(q) = [\gamma^\mu q_{\mu} - \mu + i \varepsilon]^{-1}$
being the Fermion Propagator. $\hbar q^\mu$ is the external momentum
of the electron and $\mu := m c /\hbar$ is its Compton wavevector.
If $ \gamma^\mu q_\mu$ stands on the right of  $\delta m$ it can be
replaced by $\mu$ since the external electron spinor $\psi$
fulfills the free Dirac equation $(\gamma^\mu q_\mu - \mu )\psi  
=0$. Again it is convenient to write $\delta m$ in the form $
\delta m_{vac} + \delta m_+ + \delta m_-$ where every term  
corresponds to the respective part of the photon propagator.
Then $\delta m_\pm$ can be calculated in a standard manner by
making a Feynman parametrization and performing the $k^0$
integral with the residue formula. Using $a^\mu
\gamma_{\mu} \gamma_j = - \gamma_j a^\mu \gamma_{\mu} + 2 a_j $
and $-(i/2 k) \varepsilon_{ijl} \gamma^i \gamma^j k^l k^\alpha
\gamma_{\alpha} = -k^0 \gamma_5 \gamma_l k^l / k + k \gamma_0 \gamma
_5$ as well as
\begin{equation}
  \vec{e}_\pm^i \vec{e}_\pm^{j*}  = {1\over 2} \Big \{ \delta^{ij}
  - \frac{1}{k^2} k^i k^j \mp \varepsilon^{ijl} \frac{k_l}{k}
  \Big \} \end{equation}
($\varepsilon^{123} =-1$) one arrives at
\begin{eqnarray}
\delta m_\pm &=& \frac{e^2}{(2\pi)^3 4 \varepsilon_0 c^2}
  \int_0^1 dz \int d^3 k \Big \{ \frac{1}{w^3} -\frac{1}{
  (1\mp \beta k)^2 w_\pm^3} \Big\} \times\nonumber \\ &&
  \Big\{ \pm \gamma_0 \gamma_5 k + \mp z q^0 \gamma_5\gamma_l
  \frac{k^l}{k} - (q^i+k^i)\gamma_i -z q^0 \gamma_0 +
  \frac{k^i}{k^2} \gamma_i (\vec{k}\cdot \vec{q}) \mp i
  \varepsilon^{ijl} \gamma_i q_j \frac{k_l}{k} \Big\}
\end{eqnarray}
with $w_\pm := \sqrt{z^2 \mu^2 +(\vec{k}-z\vec{q})^2+(1-z)((
k^0_\pm)^2-k^2)}$ and where $w$ is defined by the same expression
if $k^0_\pm$ is set equal to $k$. After fixing the direction of
$\vec{q}$ to be parallel to the 3-axis the angular part of the
integration can be performed as usual. In the remaining
expression a high-frequency cutoff $\kappa$ is introduced for
the $dk$ integral. As explained above this cutoff is needed
since optical activity is only present in the low-energy regime.
Since $\kappa$ has to be chosen so that $\beta \kappa \ll 1$ is
fulfilled it is sufficient to calculate $\delta m$ only to first
order
in $\beta$. In this case the remaining integrations over $z$ and
$k$ lead to closed expressions for $\delta m$ which contain
various polynomials and logarithms. For all allowed external
momenta ($q \beta \ll 1$) the lowest order Taylor expansion
\begin{equation}
\delta m_+ +\delta m_- = \frac{e^2}{9\pi^2 \varepsilon_0 c^2}
  \frac{\beta\kappa^3}{\mu^2} \Big \{ -3 \mu \gamma_0 \gamma_5
  + q \gamma_5 \gamma_3 \Big \} +O(\beta^2) + O(q^3)+O(\kappa^4)
\label{dm} \end{equation}
gives an excellent approximation to these functions.

This result deserves some comments. First, it is obvious that
$\delta m$ does not only depend on $q^\mu q_{\mu}$ as in the
vacuum but also on the three-vector $\vec{q}$. This implies that
for on-shell amplitudes with $q^\mu q_{\mu} = \mu^2$ the
propagation depends on the momentum of the electron in a
non-trivial way. The reason is that the rest frame of the medium
introduces a new time-like vector $\zeta^\mu$ so that $\delta m$
can depend on both $q^\mu q_{\mu}$ and $q^\mu \zeta_{\mu}$.

The second comment to Eq. (\ref{dm}) concerns the fact that
$\delta m_+ + \delta m_-$ depends on the spin of the electron
only through the chiral operators $\gamma_0 \gamma_5$ and
$\gamma_5 \gamma_i$ which are known to be a pseudo-scalar and
pseudo-vector, respectively. Nevertheless, $\delta m$ behaves
as a scalar under a parity transformation if one takes into
account the transformation properties of the medium. For
Fedorov's model the parameter $\beta$ is a pseudo-scalar
\cite{fedorov59,landau63} so that
$\delta m$ indeed has the right transformation properties.

The physical origin of the chiral mass corrections is hidden in
the properties of the medium. As explained in Refs.
\cite{tannoudji78,milonni83} the self energy of a particle is a
consequence of the radiation reaction and not of the vacuum
fluctuations of the electromagnetic field. Hence it is produced
by the field emanated from the electron that interacts with the
medium and turns back to the electron. The chiral nature of the
self energy is caused by the different interaction of the medium
with left and right circularly polarized light. This implies that
field modes with different total angular momentum propagate
differently (as can be seen in Eq. (\ref{dij})) and the
back reaction on the electron depends on its total angular
momentum.

To estimate the magnitude of $\delta m_+ + \delta m_-$ a specific
value for $\beta$ is needed. This can be done by applying the
method of Ref. \cite{mcclain93} to Fedorov's model. The result
is that $\beta$ is given by $\lambda (n_+ -n_-)/(4\pi)$ where
$\lambda$ is the wavelegth of the light beam and $n_\pm$ are the
refractive indices for left and right circularly polarized light.
Adopting the values for quartz (that, however, is not an {\em
isotropic} optically active medium) at a wavelength of 762 nm
one can infer that
$\beta$ is given by $ 3.6 \times 10^{-12}$ m. Then a very
large value for $\kappa$ is, e.g., $\kappa = 10^{11}$ m$^{-1}$.
Inserting these values into Eq. (\ref{dm}) gives $(\delta m_+ +
\delta m_-)/m \approx 10^{-6}$, the effect of the medium on the
electron mass is very small.
\\[2mm]
{\bf Spontaneous emission of a two-level atom}\\
Spontaneous emission of a photon during an atomic transition and
the exponential decay of the excitation probability arise from
the destructive interference between the transition amplitudes
to all photon states. The calculation of the decay factor for a
two-level atom with excited state $|e \rangle$ and ground state
$|g \rangle$ is a standard task in quantum optics and can be
found in many textbooks (see, eg. Ref. \cite{milonni94}).
In the Markov approach the
Heisenberg equations of motion of the atomic density matrix and
the annihilation operators of the radiation field is solved.
In this paper the Hamiltonian
\begin{equation} H= H_A + H_{e.m.} - \vec{d}\cdot \vec{E}^\perp
  (|e\rangle\langle g| + |g\rangle\langle e|)
\end{equation}
is used,
where $H_A := E_e |e \rangle\langle e| + E_g |g\rangle\langle g|$
contains the internal energy of the atom and $H_{e.m}$ describes
the time evolution of the free radiation field. $\vec{d} :=
\langle e | q \vec{x} | g \rangle$ is the dipole moment of the
atomic transition and $\vec{E}^\perp = - \partial_t \vec{A}^\perp$
is the transversal part of the electric field.
In  Fedorov's model $\vec{E}^\perp$ can be derived
from Eq. (\ref{lsg}) and the free evolution of the Maxwell field
is governed by the Hamiltonian
\begin{equation}
H_{e.m.} = \int d^3k \sum_{\lambda =\pm} \hbar c k^0_\lambda
  a^\dagger_\lambda(\vec{k}) a_\lambda(\vec{k})
\end{equation}
where the ground state energy was removed. Since the atom couples
only to the transversal part of the radiation field the
longitudinal and scalar part was neglected in $H_{e.m.}$.

The essential step of the Markov approach is the
Markovian approximation in the formal solution for the
annihilation operators (see, e.g., Ref. \cite{milonni94}).
Following these steps the decay rate of the excited state
is found to be
\begin{equation}
\gamma = 2\pi \int d^3 k \sum_{\lambda =\pm}|C_\lambda(\vec{k})|^2
  \delta (c k^0_\lambda - \omega_0) \; .
\label{gformel} \end{equation}
Here $\omega_0 := (E_e-E_g)/\hbar$ is the atomic transition
frequency and $C_\lambda(\vec{k})$ are the expansion coefficients
of the electric field operator defined by $\vec{d}\cdot \vec{E} =:
\hbar \int d^3k \sum_{\lambda=\pm} (i C_\lambda(\vec{k}) a_\lambda
(\vec{k}) + h.c.)$. The $\delta$ distribution guarantees energy
conservation. It also implies that the Fedorov model as well as any
other local theory of optical activity can predict the
spontaneous emission rate much better than the mass correction
since only field modes with a certain frequency can contribute
to it. The variation of the parameter $\beta$ with the field
frequency, which can only be included in nonlocal theories, does
therefore not alter the results. After evaluation of the integrals
in Eq. (\ref{gformel}) the contribution $\gamma_\pm$ of the
left and right circularly polarized modes turn out to be
\begin{equation}
\gamma_\pm = \frac{\gamma_0}{2} \frac{\hat{k}_\pm^5}{k_0^5(1\mp
  \beta \hat{k}_\pm)^2 (1\pm \beta \hat{k}_\pm)}
\end{equation}
with $k_0 := \omega_0/c$. The wavevector belonging to modes with
energy $\hbar \omega_0$ is given by $\hat{k}_\pm := k_0
(\sqrt{1+2\beta^2
k_0^2} \pm \beta k_0)/(1+\beta^2 k_0^2)$. The decay rate in free
space is denoted by $\gamma_0 := \vec{d}^2 k_0^3 /(3\pi \hbar
\varepsilon_0)$. To a very good approximation the total decay
rate $\gamma = \gamma_+ + \gamma_-$ is then given by
\begin{equation}
\gamma = \gamma_0 (1+18 \beta^2 k_0^2)
  + O(\beta^4 k_0^4) \; .
\end{equation}
This result shows that an isotropic optically active medium always
increases the spontaneous emission rate. This is due to the
fact that the decay factor, being the inverse of the lifetime,
has to transform as a scalar under parity transformations. Hence
only even powers of $\beta$ can occur in $\gamma$. That the
coefficient of $\beta^2$ is positive can be understood by looking
at $\gamma_\pm$ as a function of $\hat{k}_\pm$. Since $\hat{k}_\pm
= k_0 (1 \pm \beta k_0) + O(\beta^3 k_0^3)$ we need only to
consider the lowest order term in $\hat{k}_\pm$ depending on
$\beta$. Then $\hat{k}_+$ and $\hat{k}_-$ are shifted by the same
amount from their vacuum value, but with opposite sign. As is well
known Eq. (\ref{gformel}) states that most of the $k$-dependence
of $\gamma$ is caused by the density of the field modes
(see, e.g., Ref. \cite{milonni94}). This density increases
more if $k$ is shifted to a higher value ($\hat{k}_+$) than it
decreases if $k$ is shifted by the same amount to a lower value
($\hat{k}_-$). Thus the grow of $\gamma_+$ is larger than the
decrease of $\gamma_-$ so that the total decay rate increases.
If $\beta$ is less than zero $\gamma_-$ is larger than $\gamma_+$.
This argument also shows that the spontaneously emitted radiation
is partially polarized since for $\beta >0$, say, more
left circularly  than right circularly polarized photons are
produced.\\[2mm]
{\bf Acknowledgement}\\
It is a pleasure for me to thank J. Audretsch, R. M\"uller, and
M. Holzmann for stimulating discussions and the Deutsche
Forschungsgemeinschaft for financial support.


\begin{thebibliography}{99}
\bibitem{hillery84} M. Hillery and L.D. Mlodinow, Phys. Rev.
        A {\bf 30}, 1860 (1984).
\bibitem{drummond90} P.D. Drummond, Phys. Rev. A {\bf 42}, 6845
        (1990).
\bibitem{glauber91} R.J. Glauber and M. Lewenstein, Phys. Rev. A
        {\bf 43}, 467 (1991).
\bibitem{barnett92} S.M. Barnett, B. Huttner, and R. Loudon,
        Phys. Rev. Lett. {\bf 68}, 3698 (1992).
\bibitem{milonni95} P.W. Milonni, J. Mod. Opt. {\bf 42}, 1991
        (1995).
\bibitem{milonni94} P.W. Milonni, {\em The Quantum Vacuum},
        Academic Press, Boston 1994.
\bibitem{eimerl88} D. Eimerl, J. Opt. Soc. Am. B {\bf 5}, 1453
        (1988).
\bibitem{woolley71} R.G. Woolley, Molec. Phys. {\bf 22}, 555
        (1971).
\bibitem{fedorov59} F.I. Fedorov, Opt. Spektroskop. {\bf 6},
        49 (1959).
\bibitem{mcclain93} S.C. McClain, L.W. Hillman, and R.A. Chipman,
        J. Opt. Soc. Am. A {\bf 10}, 2383 (1993).
\bibitem{itzykson80} C. Itzykson and J.-B. Zuber, {\em Quantum
        field theory}, McGraw-Hill, New York 1980.
\bibitem{landau63} L.D. Landau and E.M. Lifshitz, {\em
        Electrodynamics of Continuous Media}, Pergamon,
        Oxford 1960.
\bibitem{tannoudji78} J. Dupont-Roc, C. Fabre, and C.
        Cohen-Tannoudji, J. Phys. B {\bf 11}, 563 (1978).
\bibitem{milonni83} P.W. Milonni, Int. J. Theor. Phys. {\bf 22},
        323 (1983).
\end{thebibliography}
\end{document}